\documentclass[useAMS,usenatbib,usegraphicx]{mn2e}

\usepackage{color,soul}

\title[On the Class II Methanol Maser Periodic Variability]{On the Class II Methanol Maser Periodic Variability due to the Rotating Spiral Shocks in the Gaps of Disks Around Young Binary Stars}

\author[]{Parfenov S. Yu.$^{1}$\thanks{E-mail:
Sergey.Parfenov@urfu.ru}, Sobolev A. M.$^1$\\
$^{1}$Ural Federal University, 51 Lenin Str., Ekaterinburg 620000, Russia\\
}

\begin{document}

\date{Accepted \today, Received \today, in original form \today}

\pagerange{\pageref{firstpage}--\pageref{lastpage}} \pubyear{2014}

\maketitle

\label{firstpage}

\begin{abstract}
We argue that the periodic variability of Class II methanol masers can be explained by variations of the dust temperature in the accretion disk around proto-binary star with at least one massive component. The dust temperature variations are caused by rotation of hot and dense material of the spiral shock wave in the disk central gap. The aim of this work is to show how different can be the Class II methanol maser brightness in the disk during the \textit{M}oment of \textit{M}aximum \textit{I}llumination by the \textit{S}piral \textit{S}hock material (hereafter MMISS) and the \textit{M}oment when the disk is \textit{I}lluminated by the \textit{S}tars \textit{O}nly (MISO).
We used the code CLOUDY (v13.02) to estimate physical conditions in the flat disk in the MISO and the MMISS. Model physical parameters of the disk were then used to estimate the brightness of 6.7, 9.9, 12.1 and 107~GHz masers at different impact parameters $p$ using LVG approximation. 
It was shown that the strong masers experience considerable brightness increase during the MMISS with respect to MISO. There can happen both flares and dips of the 107~GHz maser brightness under the MMISS conditions, depending on the properties of the system. The brightest 9.9~GHz masers in the MMISS are situated at the greater $p$ than the strong 6.7, 12.1 and 107~GHz masers that are situated at $p<200$~AU. The brightness of 9.9~GHz maser in the MMISS suppressed at $p<200$~AU and increase at $p>200$~AU.
\end{abstract}

\begin{keywords}
masers -- accretion, accretion discs
\end{keywords}

\section{Introduction}

Class II methanol masers are widespread in the regions of star formation \citep{Caswell13}. There are several hypotheses on the sites of the Class II methanol masers formation. First maser sources of this type were found toward ultra-compact HII regions \citep{Wilson84} and for some sources this association has been confirmed interferometrically (see, e.g., \citet{Menten92,Walsh98}). Indeed, Class II methanol masers can be excited by emission of the youngest HII regions with high emission measure \citep{Slysh02,Sobolev07}. Another hypothesis about association of Class II methanol masers with accretion disks was suggested by \citet{Norris98} and finds support in interferometric observations (e.g. \citet{Sugiyama14,Moscadelli14}). Actually, these two hypotheses are not in contradiction because photoevaporation of disks around massive stars can give birth to the youngest HII regions \citep{Hollenbach94}. In the case of disks the maser pumping is greatly influenced by the dust emission which has high potential to produce bright masers \citep{Sobolev97,Voronkov05}. According to the other hypothesis Class II methanol masers are formed in the outflows from the young stellar objects (e.g. \citet{DeBuizer09}). The masers in the outflows can be pumped by the heated dust. In fact it is often very difficult to discriminate between the disk and the outflow sites from the maser observations \citep{Minier00,vanderWalt07} and even by combination of the maser and non-maser observations (e.g. \citet{Edris05}) because the outflow forms at the disk surface.

Some of Class II methanol maser sources associated with massive star formation regions show periodic variability. There could be different mechanisms underlying variability of such kind (see, for example, \citet{Goedhart2005a}). It is more likely that these mechanisms have radiative nature \citep{Goedhart2005b}.

One of such mechanisms is proposed by \citet{vanderWalt11} and it relates the maser variability with background variations. Background variations itself are caused by the variability of colliding winds in a massive binary.

Other possible mechanism is related with variations of the dust infrared emission that is believed pumps Class II methanol masers \citep{Sobolev94}. As it was shown by \citet{Cragg05} the brightness temperature of Class II methanol masers could be very sensitive to variations of temperature of the dust $T_d$ that is the source of pumping infrared emission.

\citet{Inayoshi13} suggested that dust temperature variations could be caused by pulsations of a massive protostar growing under high accretion rates. \citet{Araya10} explained $T_d$ variations by the variability of the accretion rate of the disk material onto a young binary system. \citet{Goedhart2014} also claims that maser flare properties could be explained by binarity of a star associated with variable maser sources.

At the moment there are no strong evidences that the Class II methanol masers in any variable source with confirmed periodicity reside in the circumstellar disk. However, number of indications on that increases. For example, periodic variability was recently found in Cepheus A by \citet{Szymczak14}. The Class II methanol masers in this source most probably reside in the ring that is perpendicular to the jet associated with the radio continuum source HW2 \citep{Torstensson11}.

In this paper we consider possibility to explain periodic methanol maser phenomena in the model with the masers produced in accretion disk containing binary system and rotating spiral shocks in its central cavity. The structure of the circum-binary disk around low mass young binaries was studied by \citet{Ochi05, Sytov11, deCastro13}. We assume that the similar disk structure is possible in the case of a binary with massive components (see, for example, \citet{Krumholz09}).

The considered accretion disk structure has the relatively small gap in the center that is filled by the low density very hot gas and rotating spiral shocks (bow shocks) formed due to supersonic motion of binary components. The rotation of the bow shock leads to periodical changes in the radiation field within parts of the disk illuminated by the shock. This in turn leads to dust temperature variations in these parts of the disk. We used the simple disk model to estimate how the brightness of Class II methanol masers in the accretion disk could be affected by the bow shock illumination.

\section{Disk model}

\subsection{General description}
\label{section:general_description}
We consider the case when Class II methanol masers are formed in the accretion disk around the proto-binary star containing massive and intermediate mass stars. There is a gap in the disk center formed by rotating bow shocks. We assume that the gas in the accretion disk is in thermal and ionization equilibrium.

Luminosity and spectrum of the radiation in some radial direction are subject to variations due to the rotation of the bow shock. The material behind the bow shock is hot, dense and luminous in the UV and optical range, depending on the shock speed. The maximums of density and temperature of the material behind the bow shock are located in the bow shock base, i.e. the shocked region close to the surface of the massive proto-binary component. The value of gas column density, $N_{gas}$, computed along the line of sight between the disk center and the outer disk boundary (excluding the stellar material) attains its maximum when the bow shock base is on this line of sight \citep{Sytov09}. The luminosity of the radiation that strikes the inner accretion disk boundary is maximum when the boundary is illuminated by the material close to the bow shock base and this corresponds to the MMISS (Moment of Maximum Illumination by the Spiral Shock material).

The rotation of the spiral shock wave material causes relevant change of $T_d$ in the disk. This change of $T_d$ affects the pumping of methanol maser and thus leads to the maser variability.

\begin{figure}
\centering
\includegraphics[scale=0.7]{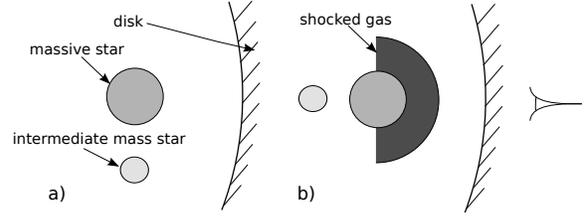}
\caption{The schematic view of two models: a) in the \textit{M}oment when the disk is \textit{I}lluminated by the \textit{S}tars \textit{O}nly (MISO); b) in the \textit{M}oment of \textit{M}aximum \textit{I}llumination by the \textit{S}piral \textit{S}hock material (MMISS). The difference of proto-binary components positions relatively to the disk inner boundary is only for clarity and was not accounted for in the calculations.}\label{ModelScheme}
\end{figure}

To estimate the maser brightness in the accretion disk under the MISO and MMISS conditions we used two models. The first model includes the disk and the proto-binary stellar system that is in the disk central gap (Fig.~\ref{ModelScheme}a). In the second model we introduce a shocked gas layer that partially fills the space between the accretion disk inner boundary and the massive component (Fig.~\ref{ModelScheme}b).
This gas layer modifies the massive star radiation before it strikes the inner disk boundary and produces great amount of radiation itself. The radiation of the intermediate mass component is not included in the second model because it is eclipsed by the massive component.

\begin{figure}
\centering
\includegraphics[scale=0.7]{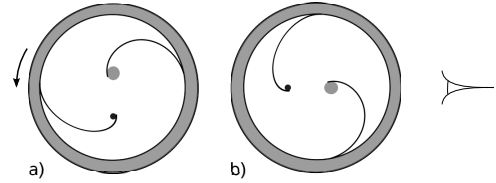}
\caption{The schematic view of proto-binary and spiral shocks configurations in the disk inner gap for two time moments which differ by the half of the binary orbital period. Filled circles show positions of proto-binary components, arcs correspond to the bow shocks, and the ring corresponds to the inner boundary of circum-binary disk. The arrow indicates the direction of binary, disk and bow shocks rotation. a) In the MISO. b) In the MMISS.}\label{ModelschemeWithShock}
\end{figure}

These two models mimic binary and spiral shocks configurations similar to those presented in Fig.~\ref{ModelschemeWithShock}.
The eclipse of the massive proto-binary component by the intermediate mass component is not considered in this study. Also we do not account for the bow shock borne by the intermediate mass component and for variations of dilution factors due to orbital motion of proto-binary components.

\subsection{Modelling with CLOUDY}

To estimate physical conditions in the circum-binary accretion disk and to compute the radiation intensity from the spiral shock material we used the photoionization code CLOUDY version C13.02 \citep{Ferland13}.
To compute physical parameters of the accretion disk under the MISO conditions we used following CLOUDY input parameters:
\begin{enumerate}
\item Central stars: the massive star and intermediate mass star. Massive star has the mass $M=13M_{\sun}$, effective temperature $T_{\rm{eff}}=29000$~K, surface gravity log$\,\textsl{g}=4.2$~dex, bolometric luminosity $L=14295L_{\sun}$, solar metallicity \citep{Grevesse98}. The atmosphere model for this star was taken from the \citet{Lanz03} grid of stellar atmosphere models. Intermediate mass star has $M=7M_{\sun}$, $T_{\rm{eff}}=20000$~K, log$\,\textsl{g}=4.2$~dex, $L=1741L_{\sun}$, solar metallicity, atmosphere model from the \citet{Castelli04} grid. Stellar luminosities and radii were obtained for the given log$\,\textsl{g}$, $T_{\rm{eff}}$ and $M$ values. Combined radiation of both stars is represented by the sum of their radiation fields from the central stellar source.
\item Geometry of the disk is flat.
\item Inner disk radius is assumed to be 1.9~AU. This radius is roughly equal to the size of the central disk gap estimated for a binary with components mass ratio of 0.54 and semi-major axis of the binary equal to 1.145~AU. The estimate was obtained by linear extrapolation of values calculated by \citet{Artymowicz94} in the framework of resonance theory.
\item Outer disk radius is 1000~AU which is close to the value obtained by \citet{Preibisch11} from their observations of the circumstellar disk around massive young stellar object in the Carina Nebula.
\item Disk semi-height is 0.38~AU. This value was obtained with the ratio of the height of the disk to its inner radius equal to 0.2, which corresponds to the estimate obtained by \citet{Preibisch11}.
\item Chemical composition of the disk is the subjective mean of the Orion Nebula abundances determined by \citet{Baldwin91, Rubin91, Osterbrock92, Rubin93}. We use the CLOUDY abundance set for HII regions.
\item We assume that the dust has CLOUDY built-in grain properties that were obtained for the Orion nebula by \citet{Baldwin91}. Grain physics described by \citet{Baldwin91} and \citet{vanHoof04}. Graphite and silicate grains distributed in 10 size bins are included in our calculations. The minimum radius of graphite and silicate grains is 0.03~$\mu$m while the maximum radius is 0.25~$\mu$m. The dust-to-gas mass ratio is 0.005. In order to simulate sublimation of the dust grains we weight dust abundance with the temperature $\exp \left[-\left(T_d/T_{sub}\right)^2\right]$, where $T_{sub}$ is the sublimation temperature for a given grain size and chemical composition. The dependence of gas-to-dust ratio on the dust temperature was chosen to provide good convergence of CLOUDY calculations.
\item Additional heating due to the viscous dissipation $H$ is treated in the framework of $\alpha$-disk model \citep{Shakura73}:
\begin{equation}
H=\alpha P\sqrt{\frac{GM}{r^3}},
\end{equation}
where $\alpha$ -- dimensionless parameter of the $\alpha$-disk model which is held constant over the disk; $P$ -- local gas pressure; $G$ -- gravitational constant; $M$ -- total mass of the binary; $r$ -- distance to the disk center.
We made calculations with $\alpha=0.00008$ and $\alpha=0.05$ which were, respectively, minimum and maximum values of $\alpha$ obtained by \citet{McClure13} by fitting spectral energy distributions of classical T~Tauri stars.
\item Hydrogen number density (the sum of number densities of ionized, neutral, molecular and other hydrogen varieties) $n_{\rm{H}}$ is proportional to inverse disk radius $r^{-1}$. This is consistent with the estimates of \citet{Ilee13} for the surface density distribution in the disks around massive stars under assumption that the surface density profile is close to the number density profile. The hydrogen number density at the inner disk edge is $10^{9.8}$~$\rm{cm}^{-3}$. Such density distribution along with the chosen stellar parameters is allows to obtain high model brightness of 6.7 and 12.1~GHz Class II methanol masers (see Section~\ref{section:maser_calculation} for details on maser brightness calculations).

Our assumption on the ionization equilibrium of the gas at the disk inner boundary is valid for the disk with the chosen parameters. This is realized due to the high hydrogen density at the inner disk edge which corresponds to the hydrogen recombination time of about tens of minutes which is small compared to the binary orbital period.
\end{enumerate}
Stellar masses and semi-major axis of the binary were chosen to obtain supersonic velocity for the orbital motion of the massive component and the binary orbital period equal to 100 days. Such period is close to observed values. 

CLOUDY input parameters to mimic physical state of the disk under the MMISS conditions are almost the same. The difference is that the disk is illuminated by combined emission of the shocked dust-free gas layer that is located in a central disk gap and the massive stellar component. This seed radiation is computed with CLOUDY assuming that the shocked gas layer has the constant density and temperature. The geometry of this layer is half of a flat disk with the inner radius and semi-height of 0.022~AU (radius of the massive star), outer radius of 0.2~AU. We made calculations for two shocked gas layer densities $n_{\rm{H}}=10^{13.62}$~$\rm{cm}^{-3}$ and $n_{\rm{H}}=10^{13.32}$~$\rm{cm}^{-3}$. Along with the outer layer radius this increases $N_{gas}$ by a factor of $\sim100$ and $\sim50$, respectively, compared to $N_{gas}$ in the model without this layer. These column density jumps are consistent with those obtained by \citet{Sytov09}. The shocked gas layer temperature is 30222~K. This temperature equals the post-shock temperature $T_s$ calculated as \citep{Lang80}:
\begin{equation}
T_s=\frac{3}{16}\frac{\mu m_{\rm{H}}v_s^2}{k_{\rm{B}}},
\end{equation}
where $m_{\rm{H}}$ -- hydrogen atom mass; $\mu$ -- mean molecular weight (for considered chemical composition it is $\sim0.7$ assuming that the gas is ionized); $k_{\rm{B}}$ -- Boltzmann constant; $v_s$ -- shock velocity equal to 43.6~km~s$^{-1}$ corresponds to the linear velocity of the massive component motion on the circular orbit. It should be noted that in reality the gas temperature behind the bow shock is not constant. The dense gas behind the shock effectively and quickly cools by intense radiation. The cooling efficiency decreases as the gas density rapidly decreases behind the bow shock. The cooling rate also decreases with decreasing gas temperature. Thus, the gas layer with temperature close to $T_s$ could be very narrow and located just immediately behind the moving bow shock. But the density and temperature of this layer are close to maximum and it should considerably affect the radiation that illuminates the inner disk boundary.

The approach of the hot shocked gas layer increases bolometric luminosity of the radiation which illuminates the inner boundary of the disk. The increase of the bolometric luminosity reaches $\sim18$ times and $\sim27$ times if the shocked gas densities are $n_{\rm{H}}=10^{13.32}$~$\rm{cm}^{-3}$ and $n_{\rm{H}}=10^{13.62}$~$\rm{cm}^{-3}$, respectively.

One can find more details of computational procedure and CLOUDY input scripts in the Appendix.

\subsection{Physical conditions in the disk}

\begin{figure}
\centering
\includegraphics[scale=0.8]{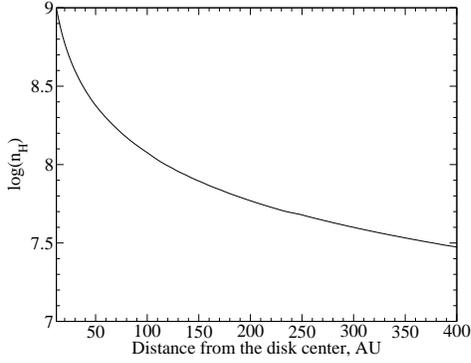}
\caption{Hydrogen number density (in cm$^{-3}$) distribution in the disk region with potential for the maser formation.}\label{nH}
\end{figure}

\begin{figure}
\centering
\center{\includegraphics[scale=1.0]{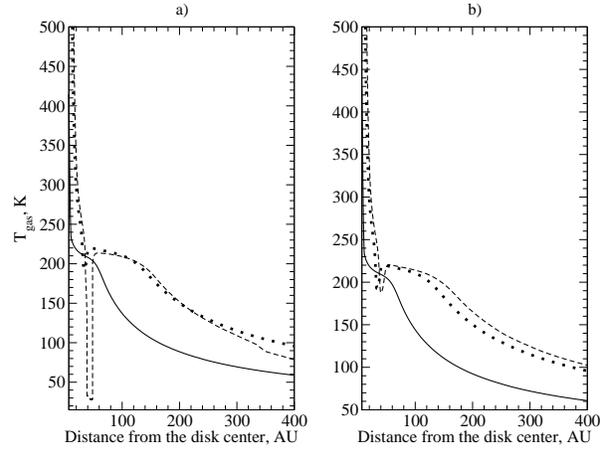}}
\caption{The gas kinetic temperature in the disk region with potential for maser formation in the MISO (\textit{solid line}) and when there is shocked gas layer in the disk center with $n_{\rm{H}}=10^{13.32}$~$\rm{cm}^{-3}$ (\textit{dotted line}) and $n_{\rm{H}}=10^{13.62}$~$\rm{cm}^{-3}$ (\textit{dashed line}). a) Disk models computed with $\alpha=0.00008$. b) Disk models computed with $\alpha=0.05$.}\label{Tgas}
\end{figure}

It follows from calculations of \citet{Cragg05} that the high brightness of Class II methanol masers could be attained only when the gas density is lower than $10^9$~$\rm{cm}^{-3}$ and the pumping dust temperature exceeds 100~K. This dust temperature is close to a limit for methanol thermal desorption \citep{Nakagawa80, Green09}. The disk region with such dust temperature and gas density is located at the range of distances from 12~AU to 120~AU from the disk center under the MISO conditions and from 12 to 400~AU under the MMISS conditions. Hereafter, we will call the latter disk region as the disk region with potential for maser formation. Hydrogen number density and gas kinetic temperature in this region calculated for models with and without shocked gas layer in the disk center are shown in Fig.~\ref{nH} and Fig.~\ref{Tgas}, respectively.

In Fig.~\ref{Tgas} it could be seen that $\alpha$ value has significant influence on the gas kinetic temperature distribution in the disk. In the case of small $\alpha$ and when shocked gas layer with $n_{\rm{H}}=10^{13.62}$~$\rm{cm}^{-3}$ presents in the disk center there is the thin disk region with relatively low gas kinetic temperature. It is caused by the thermal front where the gas changes its ionization state and transits to a different branch of the cooling curve (see CLOUDY documentation). When the the shocked gas density is equal to $n_{\rm{H}}=10^{13.32}$~$\rm{cm}^{-3}$ the thermal front is not so apparent. The temperature in this case is mostly the same as in the case of higher shocked gas density.

\begin{figure}
\centering
\center{\includegraphics[scale=1.0]{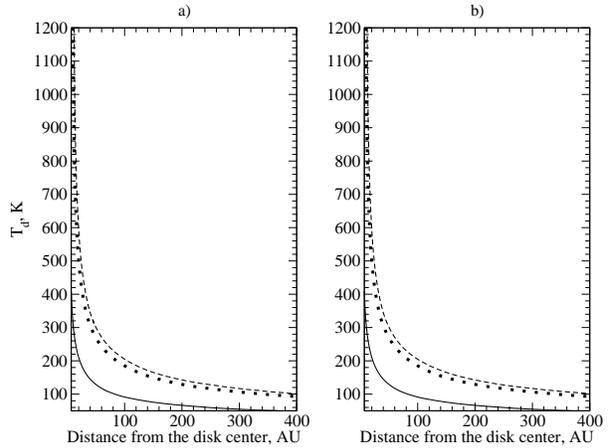}}
\caption{The temperature of small (grain radius of 0.03~$\mu$m) silicate grains in the disk region with potential for the maser formation under the MISO conditions (\textit{solid line}) and when there is shocked gas layer in the disk center with $n_{\rm{H}}=10^{13.32}$~$\rm{cm}^{-3}$ (\textit{dotted line}) and $n_{\rm{H}}=10^{13.62}$~$\rm{cm}^{-3}$ (\textit{dashed line}). a) Disk models computed with $\alpha=0.00008$. b) Disk models computed with $\alpha=0.05$.}\label{Td_sil}
\end{figure}

\begin{figure}
\centering
\center{\includegraphics[scale=1.0]{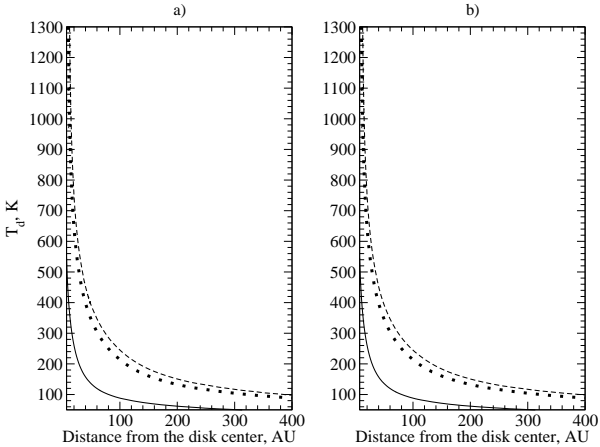}}
\caption{Same as in Fig.~\ref{Td_sil} but for small graphite grains.}\label{Td_gra}
\end{figure}

The dust temperature depends on its chemical composition and size. The temperature for smallest (grain radius of 0.03~$\mu$m) considered silicate grains in the disk region with potential for maser formation is presented in Fig.~\ref{Td_sil}. As it was shown by \citet{Ostrovskii02} a pumping dust composed of such small silicate grains is able to reproduce Class II methanol masers observed patterns. The temperature of graphite grains is presented in Fig.~\ref{Td_gra}. The temperature of the silicate and graphite grains of other sizes is somewhat greater than the shown values.

As it could be seen from Fig.~\ref{Td_sil} and Fig.~\ref{Td_gra} the illumination of the disk by the hot bow shock material leads to the increase of the dust temperature. In our model the dust in the disk region with potential for maser formation is heated mostly by diffuse emission. The dust heating rate due to diffuse emission is of order of $10^{-13}$~erg~s$^{-1}$~cm$^{-3}$. In this case the time of the dust temperature increase by 10~K is of the order of a few hours. The time of the gas temperature increase should be small compared to the binary orbital period as the thermal equilibrium between the dust and gas establishes quickly at high gas densities. Thus, our assumption on thermal equilibrium of the disk material is valid.

The dust temperature jump in the MMISS allows to preserve high methanol abundance even if the dust temperature falls below 100~K under the MISO conditions. This is because the characteristic time of methanol abundance changes greatly exceeds the binary orbital period.

The dust temperature is almost independent on the $\alpha$ value. This means that in the case of low disk viscosity there could be a disk region where both low gas kinetic temperature and high dust temperature could be realized simultaneously. Such conditions are preferable for creation of extremely bright masers because the greater the difference between gas and dust temperatures the higher the brightness of 6.7~GHz maser \citep{Cragg05}. Thus, in our model $\alpha$ is the important parameter which can greatly affect the maser brightness in the accretion disk.

\section{Modelling Class II methanol maser emission}
\label{section:maser_calculation}

\begin{figure}
\centering
\includegraphics[scale=0.8]{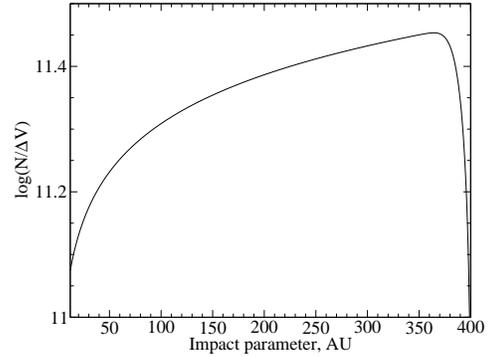}
\caption{The specific column density (in cm$^{-3}$~s) distribution.}\label{Nm}
\end{figure}

For estimated physical conditions in the accretion disk we obtained optical depths and excitation temperatures of Class II methanol masers at different impact parameters to estimate the probability of the maser action at 6.7, 9.9, 12.1 and 107~GHz. The main assumptions for these calculations were:
\begin{enumerate}
\item the disk is observed edge on;
\item physical conditions are constant along the line of sight and are the same as at the distance from the disk center that is equal to a given impact parameter.
\end{enumerate}
We used the approach from \citet{Sobolev94} to compute maser optical depths and excitation temperatures. Line transfer was treated in the large velocity gradient (LVG) approximation. Input parameters of this model were the spectrum of pumping radiation, gas kinetic temperature, hydrogen number density, beaming factor and specific column density $N_{\rm{M}}/\Delta V$. The specific column density displayed at Fig.~\ref{Nm} was obtained assuming Gaussian shape of maser lines as:
\begin{equation}
\frac{ N_{\rm{M}} }{ \Delta V }=\frac{X_{\rm{M}} f}{\Delta V \epsilon^{-1}} \int_0^{L(p)}{n_{\rm{H}}(l) \exp\left( -\frac{\left(V(p)-V(l)\right)^2}{\Delta V^2} \right)} dl,
\end{equation}
where $X_{\rm{M}}$ -- methanol abundance relative to H$_2$, it was constant and equal to $10^{-6}$; $L(p)$ -- extension of the disk region with potential for maser formation along the line of sight at the given impact parameter $p$; $f$ -- a factor that accounts for decrease of the optical depth $\tau$ due to the turbulent motions in the disk (see below); $\Delta V$ -- maser linewidth was equal to commonly observed value of 0.5~km~s$^{-1}$; $n_{\rm{H}}(l)$ -- hydrogen number density at a given distance $l$ along the line of sight; $\epsilon^{-1}$ -- beaming factor that is the ratio of the radial to tangential optical depths, it was constant and equal to 5; $V(p)$ and $V(l)$ are gas velocities at the given impact parameter and distance $l$ along the line of sight, respectively.
We assumed that the turbulent motions in the disk are of quasi-Kolmogorov type similar to those considered in \citet{Sobolev98} and \citet{Wallin98}. For our calculation we adopted the value of $f = 0.69$ which corresponds to the maximum value obtained by \citet{Sobolev98}. Gas velocities were obtained assuming Keplerian motion for the accretion disk around the binary. We considered only impact parameters exceeding 10~AU. 
The pumping radiation at considered impact parameters was computed with CLOUDY.

\begin{figure*}
\centering
\includegraphics[scale=1.0]{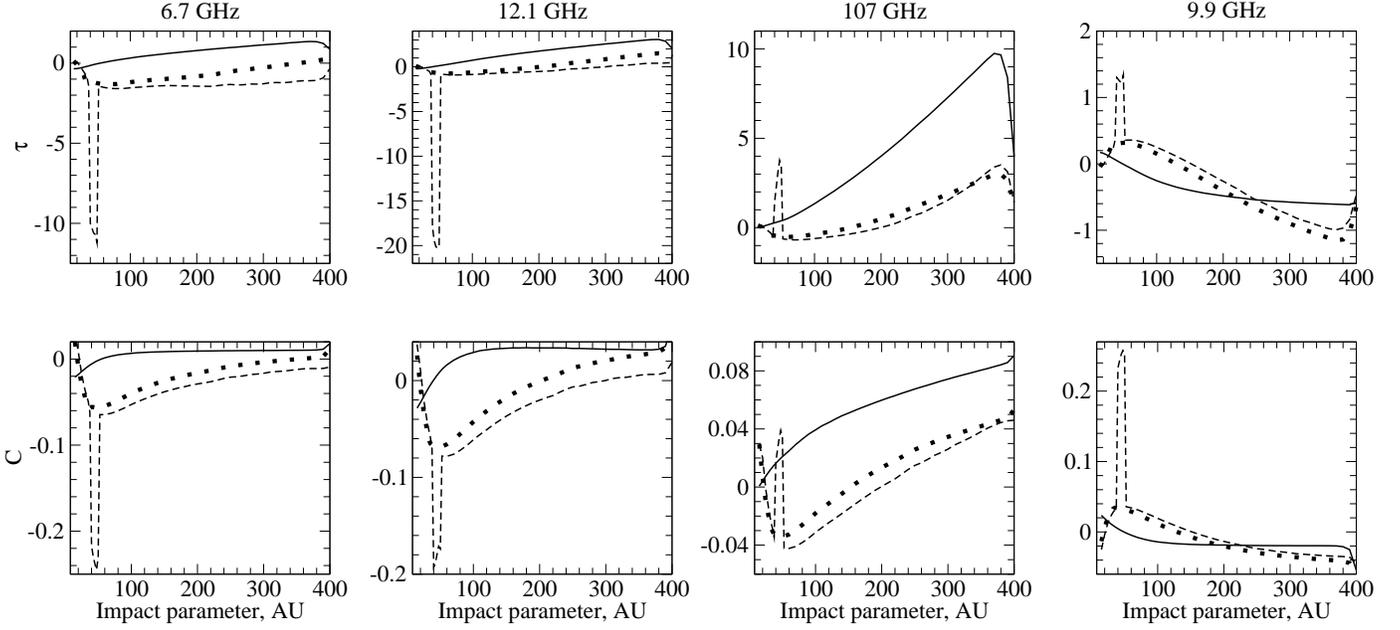}
\caption{Maser optical depths $\tau$ and excitation factors $C$ computed for disk models with $\alpha=0.00008$. \textit{Solid line} --- under the MISO conditions; \textit{dotted line} --- shocked gas layer with $n_{\rm{H}}=10^{13.32}$~$\rm{cm}^{-3}$ is located in the disk center; \textit{dashed line} --- shocked gas layer with $n_{\rm{H}}=10^{13.62}$~$\rm{cm}^{-3}$ is located in the disk center. The negative optical depth means that there is a maser amplification in a given transition while the positive value means that there is an absorption in maser transition.}\label{MaserLowVisc}
\end{figure*}

\begin{figure*}
\centering
\includegraphics[scale=1.0]{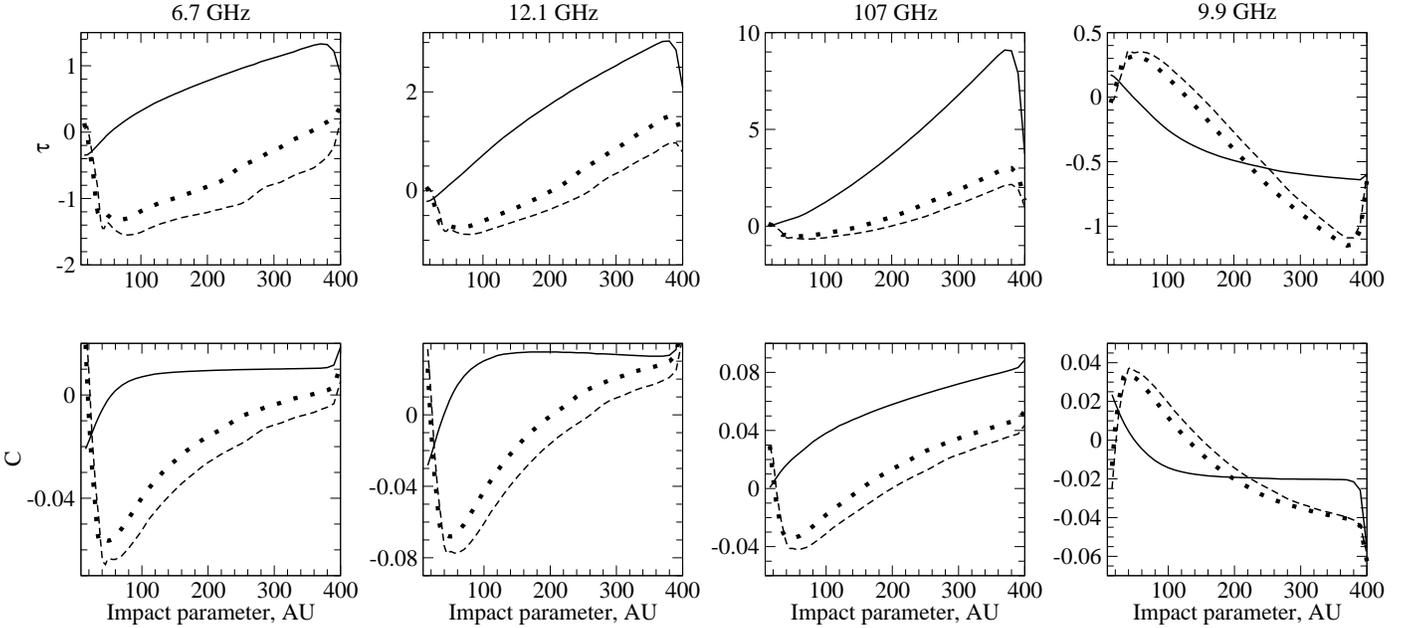}
\caption{Same as Fig.~\ref{MaserLowVisc} but for disk models computed with $\alpha=0.05$.}\label{MaserHighVisc}
\end{figure*}

Optical depths computed for $\alpha=0.00008$ and $\alpha=0.05$ are presented in Fig.~\ref{MaserLowVisc} and Fig.~\ref{MaserHighVisc}, respectively. In order to characterize the excitation of the maser transitions we present in these figures a excitation factors (originally "correction factors for stimulated emission") $C$ computed according to \citet{Goldberg66}:
\begin{equation}
C=1-\exp\left( -\frac{h\nu}{k_{\rm{B}} T_{ex}} \right),
\end{equation}
where $h$ -- Planck constant; $\nu$ -- frequency of the maser transition; $T_{ex}$ -- excitation temperature. We use the $C$ factor because it has not singularities. Negative values of the optical depth and excitation factor mean that there is a maser amplification in a given transition, while the positive values mean that there is an absorption in this transition. The decrease of the optical depth from $-1$ to $-10$ corresponds to maser brightness increase by about 4 orders of magnitude.
From Fig.~\ref{MaserLowVisc} and Fig.~\ref{MaserHighVisc} it could be seen that three disk regions could be distinguished depending on the maser brightness behavior under the MISO and MMISS conditions.

The first region of the disk is located in the range of impact parameters from 12 to 50~AU. The behavior of the maser brightness in this range of impact parameters considerably depends on the viscosity of the disk material. In this disk area under the MISO conditions the optical depth and $C$ for masers at 6.7 and 12.1~GHz are negative while $\tau$ and $C$ for 9.9~GHz transition are positive. In the MMISS $\tau$ and $C$ values for 6.7 and 12.1~GHz masers decrease while at 9.9~GHz $\tau$ and $C$ values increase. This effect greatly depends on the viscosity in the disk. When the viscosity is low the increase of $\tau$ and $C$ values for 6.7 and 12.1~GHz transitions under the MMISS conditions is much more pronounced in comparison with the case when the viscosity is high. The brightness of 6.7 and 12.1~GHz masers could increase during the MMISS event by about 4 and 8 orders of magnitude, respectively, if $\alpha = 0.00008$, and only by about 1 order of magnitude if $\alpha = 0.05$.

Within the first region of the disk $\tau$ and $C$ values for 107~GHz transition under the MISO conditions are mostly positive, i.e. the transition is quasi-thermally excited. At $p<30$~AU the 107~GHz transition is considerably overheated: $C$ decreases and approaches 0, excitation temperature exceeds 1000~K and is considerably higher than the gas kinetic temperature which ranges in this region from 200 to 400 K. If the viscosity in the disk is low the brightness of 107~GHz maser is sensitive to the density of the shocked gas layer. The brightness of 107~GHz maser can experience decrease under the MMISS conditions if the shocked gas density is high.

The second region of the disk corresponds to impact parameters of 50 to 200~AU. In this disk region $\tau$ and $C$ values for 6.7, 12.1 and 107~GHz transitions are positive under the MISO conditions and become negative in the MMISS. Values of $\tau$ and $C$ under the MISO conditions are positive for 9.9~GHz transition. In the MMISS $\tau$ and $C$ values for 9.9~GHz transition remain negative and decrease their absolute values at $150<p<200$~AU. This means that under the MMISS conditions in the range of impact parameters $150<p<200$~AU the 9.9~GHz transition remains maser which decreases its brightness. In the outer parts of this disk region the decrease of the brightness of 9.9~GHz maser becomes pronounced only when the increase of the column density $N_{gas}$ is close to its maximum value. According to \citet{Sytov09} this maximum value is about 100 and for our considered model parameters this value corresponds to the shocked gas density of $n_{\rm{H}}=10^{13.62}$~$\rm{cm}^{-3}$. In the range of impact parameters from 50 to 150~AU $\tau$ and $C$ values for 9.9~GHz transition under the MMISS conditions become positive.

The third region of the disk corresponds to impact parameters of 200--400~AU and differs from two inner regions by behavior of the maser brightness of 9.9~GHz transition. In contrast to behavior of the 9.9~GHz maser formed in the second region, $\tau$ and $C$ values for 9.9~GHz maser transition remain negative and increase their absolute values under the MMISS conditions. In our model the brightness of 9.9~GHz maser in this disk region increased by about 2 times in the MMISS.

We also made calculations for other values of the beaming factor in the range from 1 to 25. The behavior of $\tau$ and $C$ values remained the same.

\section{Conclusions}

Using simple model we have shown that in the circum-binary accretion disk with the structure similar to the one presented by \citet{Sytov11} the illumination of the disk by the bow shock hot material led to the variation of the dust temperature in the disk and considerably increased probability of the maser flare in the disk.
The influence of the disk illumination by bow shock material on the maser brightness in the disk depends on the density of the shocked material. In this study we considered the illumination by the material in the bow shock base originating close to the massive component surface. The appearance of the bow shock base in the line between considered position and the disk center corresponds to the maximum rise of column density $N_{gas}$ (given in Section~\ref{section:general_description}) and leads to considerable increase of the radiation luminosity in the considered part of the disk. This radiation luminosity decreases while the spiral shock rotates and $N_{gas}$ decreases. According to \citet{Sytov09} $N_{gas}$ rapidly increases and then slowly decays while the spiral shock rotates. We speculate that the maser brightness traces the $N_{gas}$ change with the orbital phase reproducing characteristic features of the maser flare pattern in G9.62+0.20 and G351.42+0.64, and some other sources: rapid rise and relatively slow decay \citep{Goedhart2003, Goedhart2004, vanderWalt09}. In the paper of \citet{Sytov09} it is shown that the pattern of $N_{gas}$ variation depends on inclination angle of the binary, and the maser flare pattern can change accordingly.

In our model the probability of the maser flare in the disk area close to its center depends on viscosity in the disk which is described by parameter $\alpha$. When the viscosity is relatively low ($\alpha=0.00008$) in the narrow disk area (15--50~AU from the disk center) the MMISS event can create a difference between the gas and dust temperatures up to $\sim$100 K. As a result, under the MMISS conditions the 6.7 and 12.1~GHz maser brightness in this disk area is much higher when the viscosity is low compared to the case when the viscosity is relatively high ($\alpha=0.05$) and the great temperature difference does not occur.

When the disk viscosity is low the brightness of 107~GHz maser formed at impact parameters 12--50~AU under the MMISS conditions strongly depends on the shocked gas density.  When shocked density is maximum among the considered in this study, the brightness of 107~GHz maser is lower compared to the case when the shocked gas density is lower. Thus, while the spiral shock rotates and the column density increases there could be dips of the 107~GHz maser brightness under the MMISS conditions.

The 9.9~GHz maser in our model attains its maximum brightness at greater distances from the disk center than other masers considered in this study. This means that the maser at 9.9~GHz can be observed at positions and velocities different from those of the 6.7, 12.1 and 107~GHz masers formed in the same object. Brightness of the 9.9~GHz maser formed at impact parameters 50--200~AU decreases under the MMISS conditions. The effect strongly depends on the shocked gas density and becomes less pronounced with the distance from the center, disappearing at 200~AU. 

The period of maser flares in our model is equal to the binary orbital period. If the binary consists of two massive stars then one can expect that the period of maser flares will be a half of the orbital period. The amplitude of the periodic flares is sensitive to variations in the shocked gas density and local parameters in the disk region where the masers are formed. The 6.7~GHz and 12.1~GHz transitions show almost the same dependence of the maser brightness on the illumination (MMISS or MISO) conditions. The 107~GHz and 9.9~GHz transitions can display flares or dips with the same period depending on the disk properties. 

In this paper we considered stable axisymmetric disk model with monotonous dependence of parameters on the distance from the center. Real disks are inhomogeneous. Parameters of the disks around binary systems as well as those of the binary systems are subject to time variations. This can result in the changes of variability patterns of different transitions with time. 

Episodic increases of accretion rate represent another cause of the increase of stellar luminosity and gas density in the spiral shock. This can produce an effect similar to the MMISS event. According to our calculations the variation of the luminosity by about one and a quarter order of magnitude causes considerable changes of the brightness of masers formed in the accretion disk. We did not consider changes of the accretion rate but one has to expect that similar variation of the luminosity due to episodic changes of accretion rate can lead to episodic flares/dips of masers.

\section*{Acknowledgments}

We are grateful to D.V.~Bisikalo and M.A.~Voronkov for useful comments.

\section*{Appendix}

We have changed the standard behavior of two CLOUDY commands.
The first command is "Hextra SS" and it is used to include heating due to viscous dissipation. We made so that the heating rate introduced by this command is varied not only due to gas pressure variation but also with the radius from the disk center. So, the third argument of this command became unimportant.

The second command is "save transmitted continuum". This command allows to save the sum of stellar radiation diffused through the surrounding gas and the radiation of the surrounding gas. We made so that before the summation of these two types of radiation the diffused stellar radiation is multiplied by the covering factor.

We also have added the new command "trace OCCNUM" that allowed to save photon occupation numbers at given distances from the disk center. Parameters of this command are the number of points in the disk for which occupation numbers should be saved and following list of distances from the disk center in AU.

To compute physical conditions in the disk in the MISO and with $\alpha=0.00008$ the CLOUDY input was:\\
\\
\textsl{
table star Tlusty OSTAR 3-dim temp=29000 log(g)=4.2 logZ=0.0\\
luminosity 4.155171131720616 solar\\
table star atlas odfnew 3-dim temp=20000 log(g)=4.2 logZ=0.0\\
luminosity 3.240853810488137 solar\\
cosmic rays background\\
abundances HII region no grains\\
grains Orion function\\
cylinder log semi height=12.75471480186691\\
radius 13.45368480620293 16.1749254129166\\
covering factor 0.09805806756909208\\
hden 9.8, power =-1.0\\
iterate 10\\
stop temperature 3 K linear\\
set nend 3000\\
set nchrg 5\\
age 50 days\\
set didz -3\\
set trimming -10 upper\\
set trimming -14 lower\\
Hextra SS 0.00008 20 2.78e13\\
trace OCCNUM 55 10. 15. 20. 25. 30. 32.5 35. 37.5 40. 42.5 45. 47.5 50. 52.5 55. 60. 70. 80. 90. 100. 110. 120. 130. 140. 150. 160. 170. 180. 190. 200. 210. 220. 230. 240. 250. 260. 270. 280. 290. 300. 310. 320. 330. 340. 350. 360. 370. 380. 390. 400. 420. 440. 460. 480. 500.\\
save physical conditions last "PhysCond.dat"\\
save grain temperature last "GrainTemp.dat"\\
save grain abundance last "GrainAbund.dat"\\
save grain heating last "GrainHeat.dat"\\
save element last hydr "Hion.dat"\\
save continuum last "Cont.dat"\\
save H2 temperatures last "H2Temp.dat"\\
save molecules last "Mole.dat"\\
}\\

To compute the radiation from the shocked gas layer with $n_{\rm{H}}=10^{13.62}$~$\rm{cm}^{-3}$ the CLOUDY input was: \\
\\
\textsl{
table star Tlusty OSTAR 3-dim temp=29000 log(g)=4.2 logZ=0.0\\
luminosity 4.155171131720616 solar\\
cosmic rays background\\
abundances HII region no grains\\
cylinder log semi height=11.51848788013858\\
radius 11.51848788013858 12.47595540770966\\
covering factor 0.5\\
hden 13.621\\
constant temperature, t=30222K linear\\
iterate 20\\
stop temperature 3 K linear\\
age 50 days\\
print heating\\
print ages\\
save physical conditions last "PhysCond.dat"\\
save element last hydr "Hion.dat"\\
save continuum last "Cont.dat"\\
save transmitted continuum last "radiation.txt"\\
}\\

To compute physical conditions in the disk in the MMISS, with $\alpha=0.00008$ and with shocked gas layer density of $10^{13.62}$~$\rm{cm}^{-3}$ the CLOUDY input was (comments begin with \#):\\
\\
\textsl{
table read "radiation.txt"\\
nuL(nu) = 38.71516846700204 at 0.996143 Ryd\\
\# when the shocked gas density equals to\\
\# 10$^{\wedge}$(13.32) cm$^{\wedge}$(-3)\\
\# the latter command should be replaced by\\
\# nuL(nu) = 38.54965303728508 at 0.996143 Ryd\\
cosmic rays background\\
abundances HII region no grains\\
grains Orion function\\
cylinder log semi height=12.75471480186691\\
radius 13.45368480620293 16.1749254129166\\
covering factor .2181456377060064\\
hden 9.8, power =-1.0\\
iterate 10\\
stop temperature 3 K linear\\
set nend 3000\\
set nchrg 5\\
age 50 days\\
set didz -3\\
set trimming -10 upper\\
set trimming -14 lower\\
Hextra SS 0.00008 20 2.78e13\\
print heating\\
print ages\\
trace OCCNUM 55 10. 15. 20. 25. 30. 32.5 35. 37.5 40. 42.5 45. 47.5 50. 52.5 55. 60. 70. 80. 90. 100. 110. 120. 130. 140. 150. 160. 170. 180. 190. 200. 210. 220. 230. 240. 250. 260. 270. 280. 290. 300. 310. 320. 330. 340. 350. 360. 370. 380. 390. 400. 420. 440. 460. 480. 500.\\
save physical conditions last "PhysCond.dat"\\
save grain temperature last "GrainTemp.dat"\\
save grain abundance last "GrainAbund.dat"\\
save grain heating last "GrainHeat.dat"\\
save element last hydr "Hion.dat"\\
save continuum last "Cont.dat"\\
save H2 temperatures last "H2Temp.dat"\\
save molecules last "Mole.dat"\\
}

\label{lastpage}

\end{document}